\documentclass[12pt]{article}
\usepackage{amssymb,amsfonts}
\usepackage{graphics,psboxit,amsmath}
\usepackage{subfigure}
\usepackage{graphicx}
\usepackage{verbatim}
\usepackage{color}
\usepackage{hyperref}
\hypersetup{colorlinks}

\definecolor{darkred}{rgb}{1,0,0}
\definecolor{darkgreen}{rgb}{0,0.5,0}
\definecolor{darkblue}{rgb}{0,0,1}
\definecolor{orange}{rgb}{1,0.5,0}
\definecolor{green}{rgb}{0,1,0}
\definecolor{purple}{rgb}{.5,0,1}

\hypersetup{ colorlinks,
linkcolor=darkred,
filecolor=darkgreen,
urlcolor=darkblue,
citecolor=darkblue,
linktocpage=true }

\definecolor{markcolor}{rgb}{.25,0,1}

\definecolor{markcolor2}{rgb}{1,0,0}

\definecolor{markcolor3}{rgb}{0,1,0}

\def\hybrid{\topmargin -0pt    \oddsidemargin 0.05in 
        \headheight 0pt \headsep 0pt
        \textwidth 16.0cm      
        \textheight 22,0cm       
        \marginparwidth .875in
        \parskip 5pt plus 1pt   \jot = 1.5ex}

\hybrid

\catcode`\@=11

\def\marginnote#1{}
\newcount\hour
\newcount\minute
\newtoks\amorpm
\hour=\time\divide\hour by60 \minute=\time{\multiply\hour by60
\global\advance\minute by-\hour}
\edef\standardtime{{\ifnum\hour<12 \global\amorpm={am}%
        \else\global\amorpm={pm}\advance\hour by-12 \fi
        \ifnum\hour=0 \hour=12 \fi
        \number\hour:\ifnum\minute<10 0\fi\number\minute\the\amorpm}}
\edef\militarytime{\number\hour:\ifnum\minute<10 0\fi\number\minute}

\def\draftlabel#1{{\@bsphack\if@filesw {\let\thepage\relax
   \xdef\@gtempa{\write\@auxout{\string
      \newlabel{#1}{{\@currentlabel}{\thepage}}}}}\@gtempa
   \if@nobreak \ifvmode\nobreak\fi\fi\fi\@esphack}
        \gdef\@eqnlabel{#1}}
\def\@eqnlabel{}
\def\@vacuum{}
\def\draftmarginnote#1{\marginpar{\raggedright\scriptsize\tt#1}}

\def\draft{\oddsidemargin -.5truein
        \def\@oddfoot{\sl preliminary draft \hfil
        \rm\thepage\hfil\sl\today\quad\militarytime}
        \let\@evenfoot\@oddfoot \overfullrule 3pt
        \let\label=\draftlabel
        \let\marginnote=\draftmarginnote
   \def\@eqnnum{(\theequation)\rlap{\kern\marginparsep\tt\@eqnlabel}%
\global\let\@eqnlabel\@vacuum}  }

\def\draft2{
        \def\@oddfoot{\sl preliminary draft \hfil
        \rm\thepage\hfil\sl\today\quad\militarytime}
        \let\@evenfoot\@oddfoot \overfullrule 3pt
        \let\label=\draftlabel
        \let\marginnote=\draftmarginnote
   \def\@eqnnum{(\theequation)\rlap{\kern\marginparsep\tt\@eqnlabel}%
\global\let\@eqnlabel\@vacuum}  }

\def\preprint{\twocolumn\sloppy\flushbottom\parindent 2em
        \leftmargini 2em\leftmarginv .5em\leftmarginvi .5em
        \oddsidemargin -.5in    \evensidemargin -.5in
        \columnsep .4in \footheight 0pt
        \textwidth 10.in        \topmargin  -.4in
        \headheight 12pt \topskip .4in
        \textheight 6.9in \footskip 0pt
        \def\@oddhead{\thepage\hfil\addtocounter{page}{1}\thepage}
        \let\@evenhead\@oddhead \def\@oddfoot{} \def\@evenfoot{} }

\def\numberbysection{\@addtoreset{equation}{section}
        \def\theequation{\thesection.\arabic{equation}}}

\def\underline#1{\relax\ifmmode\@@underline#1\else
        $\@@underline{\hbox{#1}}$\relax\fi}

\def\titlepage{\@restonecolfalse\if@twocolumn\@restonecoltrue\onecolumn
     \else \newpage \fi \thispagestyle{empty}\c@page\z@
        \def\thefootnote{\fnsymbol{footnote}} }

\def\endtitlepage{\if@restonecol\twocolumn \else \newpage \fi
        \def\thefootnote{\arabic{footnote}}
        \setcounter{footnote}{0}}  

\catcode`@=12 \relax

\def\figcap{\section*{Figure Captions\markboth
        {FIGURECAPTIONS}{FIGURECAPTIONS}}\list
        {Figure \arabic{enumi}:\hfill}{\settowidth\labelwidth{Figure
999:}
        \leftmargin\labelwidth
        \advance\leftmargin\labelsep\usecounter{enumi}}}
 \relax
\def\tablecap{\section*{Table Captions\markboth
        {TABLECAPTIONS}{TABLECAPTIONS}}\list
        {Table \arabic{enumi}:\hfill}{\settowidth\labelwidth{Table
999:}
        \leftmargin\labelwidth
        \advance\leftmargin\labelsep\usecounter{enumi}}}
 \relax
\def\reflist{\section*{References\markboth
        {REFLIST}{REFLIST}}\list
        {[\arabic{enumi}]\hfill}{\settowidth\labelwidth{[999]}
        \leftmargin\labelwidth
        \advance\leftmargin\labelsep\usecounter{enumi}}}
 \relax
\makeatletter
\newcounter{pubctr}
\def\publist{\@ifnextchar[{\@publist}{\@@publist}}
\def\@publist[#1]{\list
        {[\arabic{pubctr}]\hfill}{\settowidth\labelwidth{[999]}
        \leftmargin\labelwidth
        \advance\leftmargin\labelsep
        \@nmbrlisttrue\def\@listctr{pubctr}
        \setcounter{pubctr}{#1}\addtocounter{pubctr}{-1}}}
\def\@@publist{\list
        {[\arabic{pubctr}]\hfill}{\settowidth\labelwidth{[999]}
        \leftmargin\labelwidth
        \advance\leftmargin\labelsep
        \@nmbrlisttrue\def\@listctr{pubctr}}}
 \relax
\makeatother

\def\be{\begin{equation}}
\def\ee{\end{equation}}
\def\ba{\begin{eqnarray}}
\def\ea{\end{eqnarray}}

\def\p{\pi}

\def\l{\lambda}

\def\no{\noindent}

\def\IR{\relax{\rm I\kern-.18em R}}

\def\bse{\begin{small}\begin{equation*}}
\def\ese{\end{equation*}\end{small}}

\begin{document}


\renewcommand{\theequation}{\thesection.\arabic{equation}}
\csname @addtoreset\endcsname{equation}{section}

\newcommand{\eqn}[1]{(\ref{#1})}

\begin{titlepage}
\begin{center}
\strut\hfill
\vskip 1.3cm


\vskip .5in

{\Large \bf Type-I integrable quantum impurities in the Heisenberg model}

\vskip 0.5in

{\large \bf Anastasia Doikou}\phantom{x}
 \vskip 0.02in
{\footnotesize Department of Engineering Sciences, University of Patras,\\
GR-26500 Patras, Greece}
\\[2mm]
\noindent

\vskip .1cm


{\footnotesize {\tt E-mail: adoikou@upatras.gr}}\\

\end{center}

\vskip 1.0in

\centerline{\bf Abstract}
Type-I quantum impurities are investigated in the context of the integrable Heisenberg model. This type of defects is associated to the
$(q)$-harmonic oscillator algebra. The transmission matrices associated to this particular type of defects are computed via the Bethe ansatz methodology for the XXX model, as well as for the critical and non-critical  XXZ spin chain. In the attractive regime of the critical XXZ spin chain the transmission amplitudes for the breathers are also identified.

\no

\vfill

\end{titlepage}
\vfill \eject

\tableofcontents

\section{Introduction}

It has been well established by now that integrable impurities are objects that carry significant physical and algebraic meaning, hence a considerable amount of work has been devoted to their extensive study both at quantum \cite{delmusi}--\cite{doikou-transm} and classical level \cite{cozanls}--\cite{doikou-karaiskos-LL}.
The present investigation is a natural continuation of previous studies on the subject \cite{doikou-karaiskos, doikou-transm}. More precisely in \cite{doikou-karaiskos} the derivation of exact transmission amplitudes via the Bethe ansatz formulation in the context of the XXX and XXZ spin chains was established for the first time, whereas in \cite{doikou-transm} transmission matrices associated to the $\mathfrak{gl}_n,\ \mathfrak{U}_q(\mathfrak{gl}_n)$ ($n>2$) algebras were derived. Recall that the transmission matrices physically describe the interaction between the excitations of the system and the impurity, therefore they encapsulate significant physical content.
In both \cite{doikou-karaiskos, doikou-transm} the so-called type-II defects were examined; these are associated to generic representations of the $\mathfrak{gl}_n,\ \mathfrak{U}_q(\mathfrak{gl}_n)$ algebras.

Here we complete the analysis on quantum defects within the Heisenberg spin chain framework by implementing the so-called type-I defects \cite{BCZ1}. These defects as will be clear subsequently are related to the $(q)$ harmonic oscillator algebra. It is worth noting however that the transmission matrices derived in the XXZ critical regime --which may be mapped to the sine-Gordon model-- are distinct to the ones derived for the sine-Gordon in \cite{konle, corrigan}. Nevertheless, in both situations the matrices are associated to variations of the $q$-harmonic oscillator algebras, thus the overall physical factors in front of the transmission matrices turn out to be similar as will be evident in the analysis that follows.

The formulation of an one dimensional lattice integrable theory in the presence of a defect is well formulated within the context of
the quantum inverse scattering method (QISM). In general, let us consider an one dimensional $(N+1)$-site theory with a point like defect on the
$n^{th}$ site. In this case the modified monodromy matrix of the theory reads as \cite{FT}
\be
T(\lambda) = R_{0N+1}(\lambda)\ R_{0N}(\lambda) \ldots  L_{0n}(\lambda-\Theta) \ldots R_{01}(\lambda)\, ,
\label{basic0}
\ee
where $R$ corresponds to the ``bulk'' theory, $L$ corresponds to the defect,
and $\Theta$ is an arbitrary constant corresponding to the ``rapidity'' of the defect.
The Lax operator satisfies the quadratic algebra
\be
R_{12}(\lambda_1 -\lambda_2)\ L_1(\lambda_1)\ L_2(\lambda_2) = L_2(\lambda_2)\ L_1(\lambda_1)\ R_{12}(\lambda_1 -\lambda_2)\, ,
\label{basicRLL}
\ee
where the $R$-matrix is a solution of the Yang-Baxter equation (see e.g. \cite{YBE} and references therein).
The monodromy matrix of the theory $T(\l)$, naturally satisfies (\ref{basicRLL}), guaranteeing the
integrability of the model. The Hamiltonian of any generic system with a point-like defect is given e.g. in \cite{doikou-karaiskos} and may be derived as is well known via the trace of the monodromy matrix over the auxiliary space $0$.

The outline of the present work is as follows: in the next section we introduce and study the isotropic Heisenberg (XXX) spin chain in the presence of one point-like type-I defect. The relevant transmission matrices are then derived via the Bethe ansatz formulation. Similarly, in section 3 the XXZ model in the critical and non-critical regime in the presence of type-I defect is considered. In the critical regime in addition to the soliton transmission matrix the breather transmission amplitudes are also identified. A discussion on the present results as well as on possible future directions is presented in the last section.

\section{The XXX spin chain}

We start our analysis on type-I defects with the isotropic Heisenberg XXX model.
The $R$-matrix, which characterizes the bulk behavior of the monodromy matrix, is given by the Yangian \cite{yang}:
\be
R(\lambda) = \l{\mathbb I} + i{\cal P}
\ee
${\cal P} |a \rangle \otimes | b \rangle = |b \rangle \otimes |a \rangle$ is the permutation operator.
We choose to consider as the defect $L$-matrix the discrete non-linear Shcr\"{o}dinger Lax operator, (see e.g. \cite{bogo})
\be
L(\lambda) = \begin{pmatrix}
 \lambda + i {\mathbb N}+i & i a \cr
 ia^{\dag} &  i
\end{pmatrix}, ~~~~{\mathbb N} = a\ a^{\dag}. \label{L1}
\ee
The underlying algebra in this case is essentially the harmonic oscillator algebra, which naturally arises from the quadratic relation (\ref{basicRLL}), and is expressed as
\ba
&& \Big [ a,\ a^{\dag}\Big ]= 1, \cr
&& \Big [{\mathbb N},\ a \Big ] =- a, \cr
&& \Big [{\mathbb N},\ a^{\dag}\Big ] = a^{\dag}.
\ea
For our purposes here we shall also need to introduce the $L$-matrix via the crossing property
\be
\hat L_{12}(\lambda) = V_1\ L_{12}^{t_1}(-\l-i)\ V_1, ~~~~~~V_1 = \mbox{antidiag}(i,\ -i).
\ee
The derivation of the $\hat L$ matrix is necessary in order to formulate the ``unitarity" and ``crossing-unitarity" relations as will be transparent below.

The $\hat L$ matrix has the explicit form:
\be
\hat L(\lambda) = \begin{pmatrix}
  i  & -i a \cr
 -i a^{\dag} &  -\l + i{\mathbb N}
\end{pmatrix}.\label{L2}
\ee
It is easy to check that the $L$ and $\hat L$-matrices satisfy the ``unitarity property''
\be
L_{12}(\lambda)\ \hat L_{12}(-\l) = i (\lambda +i)
\ee
as well as the ``crossing-unitarity property"
\be
L^{t_{1}}_{12}(-\l-i)\ \hat L_{12}^{t_1}(\lambda-i) = -i (\lambda -i).
\ee

Our first task is to extract the respective Bethe ansatz equations. To achieve this within the algebraic
Bethe ansatz formulation we assume the existence of ``highest weight'' states locally, such that:
\ba
&& a^{\dag}\ |\omega \rangle_n = 0 ~~~~~ {\mathbb N}\ |\omega \rangle_n =0, \cr
&& \sigma^+ =\ |\omega\rangle_j =0 ~~~~~ \sigma^z\ |\omega\rangle_j = |\omega \rangle_j, ~~~~~j\neq n.
\ea
The global reference state for the model is evidently expressed as
\be
|\Omega \rangle = \otimes_{j=1}^{N+1} |\omega\rangle_j.
\ee
Then the Bethe ansatz equations (BAE) associated to the $L,\ \hat L$-matrices as defect matrices become:
\be
\mathfrak{e}^{\pm}(\lambda_i-\Theta)\ e_1^N(\lambda_i)= - \prod_{j=1}^N e_2(\lambda_i - \lambda_j), \label{BAE+}
\ee
where the plus corresponds to the $L$-matrix and the minus to the $\hat L$-matrix. $\Theta$ is the rapidity associated to the defect,
and we also define:
\be
e_k(\lambda) = {\lambda +{ik\over 2} \over \lambda - {i k \over 2}}\, ,
~~~~~~\mathfrak{e}^+(\l) = \lambda+ {i\over 2}, ~~~~~\mathfrak{e}^-(\l) ={1 \over \l - {i\over 2}}.
\ee

The process of identifying the transmission amplitudes has been described in detail in \cite{doikou-karaiskos, doikou-transm}, and the whole methodology is based on \cite{FT, Andrei-Destri}. We shall basically focus here on the derivation of the related transmission matrices.
To obtain the transmission amplitude associated to the harmonic oscillator defect it suffices to consider the state with one hole with rapidity $\tilde \l$ (low lying excitation above the ground state) \cite{doikou-nepo-mezi1}. The corresponding densities that describe the state in the thermodynamic limit are given as
\be
\sigma^{\pm}(\l) = \sigma_0(\l) +{1\over N} \Big (r^{\pm}_t(\l - \Theta) + r(\lambda - \tilde \l)\Big ). \label{sigma1}
\ee
where the $\pm$ in the densities above correspond to the $\pm$ of the BAE; also the Fourier transforms of the quantities involved in (\ref{sigma1}) are defined as
\ba
&& \hat \sigma_0(\omega) = {1\over 2 \cosh({\omega \over 2})}, \cr
&& \hat r_t^{+}(\omega) = {1\over 2 \cosh({\omega \over 2})} ~~~~\omega < 0, ~~~~~~\hat r^{+}(\omega) = 0,~~~~~~\omega > 0 \cr
&& \hat r_t^{-}(\omega) = {1\over 2 \cosh({\omega \over 2})} ~~~~\omega > 0, ~~~~~~\hat r^{-}(\omega) = 0,~~~~~~\omega < 0.
\ea
The last term in (\ref{sigma1}) corresponds to the existence of the hole in the filled Fermi sea, and provides the hole-hole scattering amplitude (see also e.g. \cite{doikou-karaiskos} and references therein).
To obtain the quantities $r_t^{\pm}$ we have also used the following Fourier transforms:
\ba
\hat a_n(\omega) &=& e^{-{n|\omega| \over 2}}, \cr
\hat {\mathfrak a}^+(\omega) &=& e^{\omega \over 2} ~~~~~\omega < 0, ~~~~~  \hat {\mathfrak a}^+(\omega) =0 ~~~~~\omega > 0 \cr
\hat {\mathfrak a}^-(\omega) &=& e^{-{\omega \over 2}} ~~~~~\omega > 0, ~~~~~  \hat {\mathfrak a}^-(\omega) =0 ~~~~~\omega < 0,
\ea
where we define:
\be
a_n(\l) ={i \over 2 \pi} {d \over d\l}\ln \Big(e_n (\lambda) \Big ), ~~~~~~~~~\mathfrak{a}^{\pm}(\l) = {i \over 2 \pi} {d \over d\l}\ln \Big(\mathfrak{e}^{\pm}(\lambda) \Big ). \label{aa}
\ee

The transmission amplitudes, which physically describe the interaction between the low lying excitations --holes in the filled Fermi sea--
and the defect are given by (see e.g. \cite{doikou-karaiskos} for more details on the derivation)
\be
T^{\pm}(\hat \lambda) = \exp\Big [ - \int_{-\infty}^{\infty} {d\omega \over \omega}\ e^{-i\omega \hat \lambda}\ \hat r_t^{\pm}(\omega) \Big ]
\label{ST}
\ee
$\hat \l = \tilde \l - \Theta$, also $T^+$ corresponds to the $L$ defect matrix, whereas $T^-$ corresponds to the $\hat L$ defect matrix.

The transmission factors may be identified through expression (\ref{ST}),  and the useful identity
\be
\int_{0}^{\infty}\ {d\omega \over \omega}\ {e^{-{\mu \omega \over 2}}\over 2 \cosh({\omega \over 2})}= \ln \Big [{\Gamma({\mu +1 \over 4}) \over \Gamma({\mu +3 \over 4})} \Big ]. \label{ident}
\ee
and are found to be
\be
T^+(\hat \lambda) =  {\Gamma(-{i\hat\lambda \over 2}+ {1\over 4})\ \over \Gamma(-{i\hat\lambda \over 2} + {3\over 4})}\, ,~~~~~~
T^-(\hat \lambda) =  {\Gamma({i\hat\lambda \over 2}+ {3\over 4})\ \over \Gamma({i\hat\lambda \over 2} + {1\over 4})}.
\label{exp1}
\ee

It is also essential to recall the $S$-matrix of the XXX model \cite{FT, Kulish-Resh},  which satisfies the Yang-Baxter equation, and is expressed as
\ba
{\mathbb S}(\lambda)  = {S_s(\lambda)\over i\lambda+1} \begin{pmatrix}
 i\lambda+1 & & & \cr
            &i\lambda  &1 & \cr
            &1 &i\lambda \ &\cr
            &&&  i\lambda+1
\end{pmatrix}, \label{Smatrix1}
\ea
where
\be
S_s(\lambda) =  {\Gamma(-{i\lambda \over 2} + {1\over 2})\ \Gamma( {i\lambda \over 2} +1) \over \Gamma(-{i\lambda \over 2} +1)\ \Gamma({i\lambda \over 2} + {1\over 2})}\, .
\ee

The transmission matrices ${\mathbb T},\ \bar {\mathbb T}$, satisfy the quadratic algebra \cite{delmusi}:
\be
{\mathbb S}_{12}(\l_1- \l_2)\ {\mathbb T}_1(\l_1)\ {\mathbb T}_2(\l_2) = {\mathbb T}_2(\l_2)\ {\mathbb T}_1(\l_1)\ {\mathbb S}_{12}(\l_1- \l_2), \label{rttb}
\ee
with the ${\mathbb S}$-matrix defined above (\ref{Smatrix1}). Given that, one may check that the transmission matrices may be cast as
\ba
{\mathbb T}(\hat \l)= {T^-(\hat \l)\over i\l +{1\over 2}} \begin{pmatrix}
 i\hat \lambda+1 + \bar {\mathbb N} & a \cr
 a^{\dag}                & 1 \
\end{pmatrix},  ~~~~~~~~~\bar {\mathbb N}= a\ a ^{\dag} - {1\over 2}\label{Tmatrix1}
\ea
whereas the conjugate transmission matrix is given as:
\ba
\bar {\mathbb T}(\hat \l)= T^+(\hat \l)\ \begin{pmatrix}
1   & -a \cr
 -a^{\dag}                & -i\hat \l + \bar {\mathbb N} \
\end{pmatrix}. \label{Tmatrix2}
\ea
Note that the operator $\bar {\mathbb N}$  is shifted by a factor ${1\over 2}$, compared to  the operator ${\mathbb N}$ of the ``bare" defect matrix $L$ ($\hat L$) as is also verified by the computations of the quantum numbers via the Bethe ansatz equations, and the asymptotic behavior of the transfer matrix.

It is worth noting here that both ${\mathbb T}$ and $\bar {\mathbb T}$ matrices satisfy the quadratic algebra (\ref{rttb}),
therefore are somehow equivalent algebraic entities.
The overall factors in front of each expression
are compatible with the BAE, as well as the unitarity and the crossing property requirements, i.e.
\ba
&& {\mathbb T}_{12}(\hat \l)\ \bar {\mathbb T}_{12}(-\hat  \l) = {\mathbb I}, \cr
&& \bar {\mathbb T}_{12}^{t_1}(\hat \l + i)\ {\mathbb T}_{12}^{t_1}(-\hat \l+i) ={\mathbb I}. \label{uni-cross}
\ea
The latter relations have been explicitly checked and verified for the transmission matrices derived above. Thus the validity of the transmission amplitudes derived thought the BAE is further confirmed via (\ref{uni-cross}), this of course provides an extra validity check.

\section{The XXZ spin chain}

We now proceed with the analysis of the type-I defects within the XXZ spin chain.
Let us first introduce the $R$-matrix associated to the XXZ model, which has the familiar form \cite{kulres}
\be
R(\lambda) = \begin{pmatrix}
 e^{\mu\lambda}q^{{1\over 2} +{\sigma^z \over 2}} -e^{-\mu\lambda}q^{-{1\over 2}-{\sigma^z \over 2}} &  (q-q^{-1})\ \sigma^- \cr
 (q-q^{-1})\ \sigma^+ &  e^{\mu\lambda}q^{{1\over 2} -{\sigma^z \over 2}} -e^{-\mu\lambda}q^{-{1\over 2}+{\sigma^z \over 2}}
\end{pmatrix},
\ee
$q=e^{i\mu}$. In the critical regime (no mass gap) $\mu$ is real, whereas in the non critical case (mass gap) $\mu =i\eta, ~ q=e^{-\eta}$, $\ \eta$ is real.

The defect $L$-matrix we choose to consider here is associated to the $q$-deformed oscillator algebra:
\ba
L(\lambda) =  \begin{pmatrix}
 e^{\mu\lambda}q^{1\over 2}V -e^{-\mu\lambda}q^{-{1\over 2}}V^{-1}  &  a^{\dag} \cr
 a &  -e^{-\mu\lambda}q^{-{1\over 2}}V
\end{pmatrix}. \label{L2Z}
\ea
The latter matrix corresponds to the so-called Discrete-Self-Trapping (DST) model \cite{dst},
and may be thought of as a $q$-deformation of the discrete non-linear Schrodinger model.
It is convenient to parameterize the $q$-oscillator algebra via the Weyl-Heisenberg elements:
\be
{\mathbb X}\ {\mathbb Y} = q\ {\mathbb Y}\ {\mathbb X}.
\ee
More precisely,
\be
V= {\mathbb X}, ~~~~~a^{\dag} = ({\mathbb X}^{-1} - q {\mathbb X}){\mathbb Y}^{-1}, ~~~~~~a= {\mathbb Y}\ {\mathbb X}.
\ee
The algebra associated to the $L$-matrix above as already mentioned is the $q$-harmonic algebra, and is expressed as
\ba
&& a^{\dag}\ a= 1 -q V^2, \cr
&&a\ a^{\dag} = 1 - q^{-1}V^2\cr
&& V\ a = q\ a\ V \cr
&&V\ a^{\dag} = q^{-1} a^{\dag}\ V.
\ea
The conjugate $\hat L$ matrix, obtained via the crossing property
\be
\hat L_{12}(\lambda) = V_1\ L_{12}^{t_1}(-\l-i)\ V_1, ~~~~~~V_1 = \mbox{antidiag}(i,\ -i).
\ee
is then given as
\ba
\hat L(\lambda) =  \begin{pmatrix}
- e^{\mu\lambda}q^{1\over 2} V  &  -a^{\dag} \cr
- a &  e^{-\mu\lambda}q^{-{1\over 2}}V-e^{\mu\lambda}q^{{1\over 2}}V^{-1}
\end{pmatrix}. \label{L2b}
\ea
The $L,\ \hat L$ matrices satisfy the unitarity and crossing unitarity properties as:
\ba
&& L_{12}(\l)\ \hat L_{12}(-\l) = - e^{-\mu \l}(e^{\mu\l}- e^{-\mu \l}) \cr
&& L_{12}^{t_1}(-\l -i)\ \hat  L_{12}^{t_1}(\l -i) = e^{\mu \l}(e^{\mu\l}- e^{-\mu \l}).
\ea

As in the previous section in order to derive the Bethe ansatz equations we consider highest (lowest) weight states.
It is convenient here, due to the parametrization of the $q$-harmonic algebra to consider ``lowest weight'' states
as local reference states:
\ba
&& a^{\dag}\ |\omega \rangle_n =0 ~~~~~~V\ |\omega \rangle_n = q^{1 \over 2}\ |\omega \rangle_n \cr
&& \sigma^-\ |\omega \rangle_j = 0  ~~~~~~~\sigma^z\ |\omega\rangle_j= - |\omega\rangle_j, ~~~~~~j\neq n.
\ea
The global reference state then is:
\be
|\Omega\rangle = \otimes_{j=1}^{N+1} |\omega\rangle_j.
\ee
The Bethe ansatz equations associated to the $L$ and $\hat L$ matrices are given by expressions (\ref{BAE+}), where we define in the critical XXZ case:
\ba
&& e_n(\lambda) = {\sinh(\mu(\lambda + {in\over 2})) \over \sinh((\mu(\lambda - {in\over2}))}
 \cr
&& {\mathfrak e}^+(\lambda) = {e^{-\mu\l} \over \sinh(\mu(\l + {i\over 2}))} \cr
&& {\mathfrak e}^-(\l) = e^{-\mu\l} \sinh(\mu (\l - {i\over 2})), \label{crit}
\ea
and in the non-critical regime:
\ba
&& e_n(\lambda) = {\sin(\eta(\lambda + {in\over 2})) \over \sin(\eta(\lambda - {in\over2}))}
 \cr
&& {\mathfrak e}^+(\lambda) = {e^{-i\eta\l} \over \sin(\eta(\l + {i\over 2}))} \cr
&& {\mathfrak e}^-(\l) = e^{-i\eta\l} \sin(\eta(\l - {i\over 2})). \label{crit}
\ea

\subsection{The critical case}

We shall distinguish two regimes in the critical case that is the repulsive and the attractive one depending on the values of the anisotropy parameter. This regime may be suitably mapped to the sine-Gordon model. Therefore, it is useful to provide the
relation between the sine-Gordon coupling constant $\beta^2$, and the anisotropy parameter $\mu$ of the XXZ model (see also e.g. \cite{doikou-nepo-breather} for a more detailed discussion):
\ba
\beta^2 &=& 8(\pi - \mu)\, ,~~~~~4\pi < \beta^2 < 8 \p ~~~~~~~~~ \mbox{repulsive regime}, \cr
\beta^2 &=& 8\mu\, ,  ~~~~~~~~~~~~~~0< \beta^2 < 4\pi ~~~~~~~~~~\mbox{attractive regime}.
\ea
Note that in the attractive regime the formulation of bounds states between solitons
and anti-solitons of zero spin (topological charge), the so called ``breathers'' is
allowed.

The main focus here will be in the attractive regime; basically due to the existence of  bound states.
Thus the scattering between the breathers and the defect may be also investigated in this context.
Analogous results may be extracted in the repulsive regime (see e.g. \cite{doikou-karaiskos}), which will not be treated here for brevity.
As was shown in earlier studies, the ground state in the attractive regime consists of the so-called
negative parity strings (see also e.g. \cite{doikou-nepo-breather} and references therein)
\be
\lambda^{(-)} = \lambda + {i\pi \over 2 \mu}\, .
\ee
The BAE are then modified as follows, compared to (\ref{BAE+}),
\be
{\mathfrak g}^{\pm}(\lambda_i-\Theta)\ g_1^N(\lambda) = - \prod_{j=1}^M e_2(\lambda_i - \lambda_j)\, ,
\ee
where we define
\ba
&& g_n(\lambda)={\cosh (\mu(\lambda +{in\over 2})) \over \cosh(\mu (\lambda - {in \over 2}))}\, \cr
&& {\mathfrak g}^{+}(\l) = {e^{-\mu\lambda} \over \cosh(\mu(\l + {i\over 2}))}, ~~~~~
{\mathfrak g}^{-}(\l) = e^{-\mu\lambda} \cosh(\mu(\l - {i\over 2})).
\ea

A generic state with one particle excitation of rapidity $\tilde \l$ (one hole in the filled Fermi sea of
negative parity strings) is considered and the density associated to this state may be
derived based on the standard formulation \cite{FT, Andrei-Destri}. It turns out that the derived state
density is given by the following expression
\be
-\sigma^{\pm}(\lambda) = b_1(\lambda) + {1\over N} {\mathfrak b}^{\pm}(\lambda-\Theta) - \int_{-\infty}^{\infty}\ d\lambda'\ a_2(\lambda -\lambda')\ \sigma(\lambda') + {1\over N} a_2(\lambda - \tilde \lambda) \, . \label{ss2}
\ee
where
\be
b_n(\l) = {i\over 2 \pi} {d\over d\l} \ln \Big ( g_n(\l) \Big ), ~~~~~
{\mathfrak b}^{\pm} (\l)= {i\over 2 \pi} {d \over d\l} \ln \Big ({\mathfrak g}^{\pm}(\l)\Big ).
\ee
The Fourier transformations of the quantities appearing in (\ref{ss2}) are found to be
\ba
\hat a_n(\omega) &=&  {\sinh ((\nu -n ){ \omega \over 2}) \over \sinh  ({\nu \omega \over 2})}\, , ~~~0 < n < 2 \nu \, , \cr
\hat b_n(\omega) &=& - {\sinh (  {n \omega \over 2}) \over \sinh  ({\nu \omega \over 2})}\, , ~~~0 < n < \nu\, , \cr
\hat b_n(\omega) &=& - {\sinh ( (n- 2\nu) {\omega \over 2}) \over \sinh  ({\nu \omega \over 2})}\, , ~~~\nu <n < 2\nu\, , \cr
\hat {\mathfrak b}^{\pm}(\omega) &=& \pm {e^{\pm {\omega \over 2}} \over 2 \sinh ({\nu \omega \over 2})}.
\label{bfour}
\ea
Similarly to the previous
section, the density is compactly expressed as
\be
\sigma^{\pm}(\lambda) = \sigma_0(\lambda) + {1\over N} \Big (r(\lambda-\tilde \lambda) + r^{\pm}_t(\lambda-\Theta) \Big )\, , \label{dens3}
\ee
whereas the Fourier transforms of the latter quantities are given by
\ba
&&\hat \sigma_0(\omega)  = {1\over 2 \cosh ((\nu -1){\omega \over 2})}\,,
\cr
&& \hat r_t^{\pm}(\omega) = {\mp} {e^{\pm{\omega \over 2}}\over 4\sinh({\omega \over 2}) \cosh ((\nu-1){\omega \over 2})}\, .
\ea
The quantity $r$  provides the soliton-soliton scattering amplitude (see e.g. \cite{doikou-karaiskos} and references therein for more details).

The $S$-matrix in this regime, solution of the Yang-Baxter equation as well, is given as \cite{zamo}
\ba
{\mathbb S}(\lambda)  = {S_s(\lambda, \gamma) \over a(\lambda, \gamma)}\begin{pmatrix}
 a(\lambda,\gamma)  & & & \cr
            &b(\lambda,\gamma)&c(\gamma)& \cr
            &c(\gamma)&b(\lambda, \gamma)&\cr
            &&&a(\lambda,\gamma)
\end{pmatrix} \, ,
\label{Smatrix2}
\ea
where we define $\gamma = \nu-1$
\ba
S_s(\lambda, \gamma) & = & \prod_{k=0}^{\infty}
{\Gamma(i\l +2(k+1)\gamma)\ \Gamma(i\l + 2k \gamma +1) \over \Gamma(i\l +(2k+1)\gamma)\ \Gamma(i\l + (2k +1)\gamma +1)}  \cr
& \times & \,
{\Gamma(-i\l +(2k+1)\gamma)\ \Gamma(-i\l + (2k+1) \gamma +1) \over \Gamma(-i\l +2(k+1)\gamma)\ \Gamma(-i\l + 2k \gamma +1) } \, ,
\label{S2}
\ea
and
\be
a(\lambda, \gamma) = \sin(\pi(i\lambda +  \gamma))\, ,  ~~~~\beta(\lambda,\gamma) =\sin(i\pi\lambda)\, ,
~~~~c(\gamma) = \sin(\pi \gamma)\, .
\ee

The $T^{\pm}$ amplitudes are derived from BAE and recalling (\ref{ST}) as well as the following identity
\be
{1 \over 4} \int_{-\infty}^{\infty}\ {dx \over x}\ {e^{-{\mathrm m}x}\over \sinh(x)\ \sinh (\beta x)} = \ln \prod_{k=0}^{\infty} \Gamma({{\mathrm m}\over 2} + {\beta \over 2} + k \beta + {1\over 2}).
\ee
we conclude that:
\ba
&& T^{+}(\hat \l, \gamma)  = \prod_{k=0}^{\infty}
{\Gamma(i \hat \l  +{\gamma \over 2}+2k\gamma)\ \Gamma(-i \hat \l  +{\gamma \over 2} +2k\gamma+1) \over \Gamma(i \hat \l  +{\gamma \over 2}+(2k+1)\gamma)\ \Gamma(- i \hat \l + {\gamma \over 2}+(2k+1)\gamma+1) }\cr
&&T^-(\hat \lambda, \gamma) =  \prod_{k=0}^{\infty}
{\Gamma(i \hat \l  +{\gamma \over 2}+(2k+1)\gamma+1)\ \Gamma(-i \hat \l  +{\gamma \over 2} +(2k+1)\gamma)\over \Gamma(i \hat \l  +{\gamma \over 2}+2k\gamma+1)\ \Gamma(-i \hat \l  + {\gamma \over 2}+2k\gamma) }.
\label{expb2}
\ea
The transmission matrices ${\mathbb T},\ \bar {\mathbb T}$ satisfy the quadratic algebra (\ref{rttb})
together with the $S$-matrix presented above (\ref{S2}), thus
we conclude that the transmission matrices can be expressed as
\ba
{\mathbb T}(\hat \l, \gamma)= {e^{-{\tilde \mu u \over 2}}\ T^{-}(\hat \l, \gamma)\over e^{-\tilde \mu u}\tilde q^{{1\over 2}}-e^{\tilde \mu u}\tilde q^{-{1\over 2}} }\begin{pmatrix}
 e^{-\tilde\mu u}\tilde q V -e^{\tilde\mu u}\tilde q^{-1}V^{-1}  &  a^{\dag} \cr
 a &  -e^{\tilde \mu u}\tilde q^{-1}V
\end{pmatrix}, \
\ea
and
\ba
\bar {\mathbb T}(\hat \lambda, \gamma) = - e^{{\tilde \mu u\over 2}}\ \tilde q^{1\over 2}\ T^+(\hat  \l, \gamma)  \begin{pmatrix}
- e^{-\tilde \mu u} V  &  -a^{\dag} \cr
- a &  e^{\tilde \mu u} V - e^{-\tilde \mu u}V^{-1}
\end{pmatrix}, \label{T2b}
\ea
where
\be
u = {\hat \l \over \gamma}, ~~~~~\tilde \mu = \pi \gamma, ~~~~~\tilde q = e^{i\tilde \mu}.
\ee
Crossing and unitarity properties (\ref{uni-cross}) ($\hat \l \to u$) are satisfied by the matrices ${\mathbb T},\ \bar {\mathbb T}$  as has been explicitly verified.

We may compare the transmission amplitudes with earlier results on type-I defects \cite{konle, corrigan}, although it is worth mentioning that the transmission matrices, as well as the underlying algebras are quite different. Nevertheless, both algebras are essentially variations of the $q$-deformed oscillator, and therefore the derived transmission amplitudes turn out to be similar. Indeed, consider for instance the expression for $T^+$ (\ref{expb2}), by setting ${\mathrm z} \equiv i\hat \l - {\gamma \over 2} - {1\over 2}$ one recovers the transmission amplitude derived in \cite{konle, corrigan}.

\subsubsection*{Breathers}
Having identified the soliton transmission matrix we may now proceed with the derivation of the breather type-I transmission amplitude. The breathers are in general identified within the Bethe ansatz frame by suitable ``string configurations'' via the so-called string hypothesis \cite{FT}. A state with two light breathers with rapidities $\bar \lambda_1,\ \bar \lambda_2$ will be considered. We shall basically deal with the lightest breathers for simplicity; a generalization of the results concerning higher breathers is then straightforward \cite{doikou-nepo-breather}, and is given at the end of the subsection. The lightest breather is described by one positive parity (real) string with rapidity $\bar \lambda_j$, then the BAE for the state with two light breathers are expressed as :
\be
{\mathfrak g}^{\pm}(\lambda_i -\Theta)\ g_1^N(\lambda_i) = - \prod_{j=1}^M e_2(\lambda_i -\lambda_j)\ \prod_{j=1}^2 g_{2}(\lambda_i - \bar \lambda_j). \label{set1}
\ee
To describe and understand the scattering process for the breathers it is necessary to take into consideration two sets of Bethe ansatz equations; the first set describes the negative parity one-strings, while the second one describes the breather itself. The second set of BAE is necessary in order to derive the energy and momentum of the breather, and also compare with the quantization condition with respect to the breather (for more details see e.g \cite{doikou-nepo-breather} and references therein). The second set of BAE describing the breather with rapidity $\bar \lambda_1$ is derived as
\be
{\mathfrak e}^{\pm}(\bar \lambda_1 -\Theta)\ e_1^N(\bar \lambda_1) = - \prod_{j=1}^M g_2(\bar \lambda_1 -\lambda_j)\ e_{2}(\bar \lambda_1 - \bar \lambda_2).
\label{set2}
\ee
We shall also need in what follows the Fourier transform of ${\mathfrak a}^{\pm}$ (\ref{aa}) derived as:
\be
\hat {\mathfrak a}^+(\omega) ={ e^{-(\nu-1){\omega \over 2}} \over 2 \sinh({\nu \omega \over 2})}, ~~~~~~~\hat {\mathfrak a}^-(\omega) =
-{ e^{(\nu-1){\omega \over 2}} \over 2 \sinh({\nu \omega \over 2})}.
\ee

From the first set of BAE (\ref{set1}) the following density regarding the negative parity strings arises,
\be
\sigma^{\pm}(\lambda) = \sigma_0(\lambda) + {1\over N} \Big ({\mathrm B}^{\pm}(\lambda-\Theta) + \sum_{j=1}^2{\mathrm R}(\lambda -\tilde \lambda_j) \Big ),
\ee
where we define the Fourier transforms of ${\mathrm B}^{\pm}$ as
\be
\hat {\mathrm B}^{\pm}(\omega) = \mp {e^{\pm{\omega\over 2}} \over 4 \sinh ({\omega \over 2})\cosh((\nu-1){\omega \over 2})}
\ee
${\mathrm R}$ is associated to the breather scattering amplitude (see e.g. \cite{doikou-karaiskos, doikou-nepo-breather} for more details).

The second set (\ref{set2}) leads to the density describing the breather
\be
\bar \sigma^{\pm}(\lambda) = \bar \sigma_0(\lambda) + \Big ( t^{\pm}_b(\lambda-\Theta) + \sum_{j=1}^2 r_b(\lambda - \bar \lambda_j) \Big ),
\ee
where the corresponding Fourier transforms read as
\be
\hat {\bar \sigma}_0(\omega) ={\cosh((\nu-2){\omega \over 2})\over \cosh((\nu-1){\omega \over 2})},
~~~~~\hat t^{\pm}_b(\omega)= -{e^{\mp(\nu-2){\omega \over 2}} \over 2 \cosh((\nu-1){\omega \over 2})}. \label{four4}
\ee
$r_b$ gives rise to the breather scattering amplitude \cite{doikou-karaiskos}.

The expressions for the breather transmission amplitude are given as (see also \cite{doikou-karaiskos, doikou-nepo-breather}):
\be
T_b^{\pm(1)}(\hat \lambda, \gamma) = \exp \Big [ - \int_{-\infty}^{\infty}\ {d\omega \over \omega} e^{-i \omega \hat \lambda} \hat t^{\pm}_b(\omega)\Big ] \label{exp4}
\ee
$\hat \lambda = \bar \lambda_1 -\Theta$.

Through (\ref{exp4}) and (\ref{ident}) the breather transmission amplitudes may be identified as
\ba
T_b^{+(1)}(\hat \theta, \gamma) =- {\sinh({\hat \theta \over 2} +{i\pi \over 4 \gamma})
\over \sinh({\hat \theta \over 2} +{i\pi \over 4\gamma } - {i\pi \over 2})} \cr
T_b^{-(1)}(\hat \theta, \gamma) =-{\sinh({\hat \theta \over 2} -{i\pi \over 4 \gamma}- {i\pi \over 2})
\over \sinh({\hat \theta \over 2} -{i\pi \over 4\gamma })}
\label{breatherfinal}
\ea
$\hat \theta = {\pi \hat \lambda \over \gamma}$, and it is clear that our expressions (\ref{breatherfinal}) for the breather transmission amplitudes coincide with those identified in \cite{corrigan} up to a shift in the spectral parameter. Notice that crossing is also verified between the two latter amplitudes that is:
$T^{-(1)}_b(\hat \theta, \gamma) = T^{+(1)}_b(-\hat \theta +i\pi, \gamma)$.

The results on the transmission amplitudes may be generalized for higher $n$-breathers, which are represented by
$n$-positive parity strings with real centers $\bar \lambda_j$ (see also \cite{doikou-karaiskos}). More precisely, it is straightforward to see (see e.g. \cite{doikou-nepo-breather}) the transmission amplitude of an $n$-breather is
\be
T_b^{(n)}(\hat \lambda, \gamma)= \prod_{l=1}^{n}\ T_b^{(1)}\Big (\hat \lambda + {i\over 2} (n +1 -2 l), \gamma\Big ).
\ee
This concludes our analysis in the critical regime of the XXZ model.

\subsection{The non-critical case}

We shall investigate in this section both type-I and II transmission matrices in the non-critical regime of the XXZ model. Before we proceed with our analysis let us first introduce some useful technical points. In this regime we basically deal with discrete Fourier transforms, and the relevant conventions are set as follows (see e.g. \cite{doikou-nepo-mezi2} and references therein):
\be
f(\l) = {\eta \over \pi} \sum_{k=-\infty}^{\infty} e^{-2i\eta k\l}\hat f(k), ~~~~~
\hat f(k) = \int_{-{\pi \over 2\eta}}^{{\pi \over 2 \eta}}\ d\l\ e^{2i \eta k\l} f(\l).
\ee
Using the latter conventions we derive the following useful Fourier transforms:
\ba
&& \hat a_n(k) = e^{-n \eta |k|} \cr
&& \hat {\mathfrak a}^{+}(k) = -e^{\eta k} ~~~~k<0, ~~~~~\hat {\mathfrak a}^+(k) =0 ~~~~k > 0 \cr
&& \hat {\mathfrak a}^{-}(k) = -e^{-\eta k} ~~~~k>0, ~~~~~\hat {\mathfrak a}^-(k) =0 ~~~~k < 0.
\ea
In this regime the main modification compared to the critical case is that the hyperbolic functions turn to trigonometric ones. As in the cases examined in the previous sections in order to derive the transmission amplitudes it suffices to consider a state with one hole of rapidity $\tilde \l$ in the filled Fermi sea. Then the density of the state is given by (\ref{dens3}), and the Fourier transform of the quantities appearing in (\ref{dens3}) are defined as follows
\ba
&& \hat \sigma_0(\omega) = {1\over 2 \cosh(\eta k)}, \cr
&& \hat r_t^{+}(\omega) =-{1\over 2 \cosh(\eta k)} ~~~~~ k < 0, ~~~~~\hat r_t^{+}(\omega) =0 ~~~~k>0\cr
&& \hat r_t^{-}(\omega) =-{1\over 2 \cosh(\eta k)} ~~~~~ k > 0, ~~~~~\hat r_t^{-}(\omega) =0 ~~~~k<0.
\ea

It is also a good opportunity at this point to present the bulk $S$-matrix derivation in the non critical regime of the XXZ model.
As explained in detail in previous works \cite{FT}, to obtain the bulk $S$-matrix one needs to deal with a state consisting of two holes with rapidities $\tilde \l_1,\ \tilde \l_2$  in the filled Fermi sea. Then the hole-hole scattering amplitude of the non critical XXZ model is given by:
\ba
&& S(\l) =  \exp \Big [ - \sum_{k=-\infty}^{\infty}  {1\over k}\ e^{-2i\eta k\l}\ \hat r(k) \Big ] \cr
&& \hat r(k) = {e^{-2 \eta |k|} \over  1+e^{-2 \eta |k|}}.
\ea
$\l = \tilde \l_1- \tilde \l_2$.
Recall also the useful identity (see \cite{doikou-nepo-mezi2} and references therein)
\be
\sum_{k=1}^{\infty}\ {1\over k} { e^{-2\eta k x} \over 1 + e^{-2 \eta k} }= \ln \Big [{\Gamma_{q^4}({x\over 2 }) \over \Gamma_{q^4}({x\over 2 } +{1\over 2}) } \Big ] - {1\over 2} \ln(1-q^4). \label{identnc}
\ee
The $\Gamma_q$-function is the $q$ analogue of the familiar Euler $\Gamma$-function, and is defined as:
\be
\Gamma_q(x) = (1-q)^{1-x} \prod_{j=0}^{\infty} {1-q^{1+j} \over 1-q^{x+j}}
\ee
recall $q=e^{-\eta}$. We may then express the scattering amplitude above in terms of $\Gamma_q$-functions as
\be
S_s(\lambda) = {\Gamma_{q^4}(-{i\l \over 2}+{1\over 2}) \over \Gamma_{q^4}(-{i\l \over 2}+1)}\ {\Gamma_{q^4}({i\l \over 2}+1) \over \Gamma_{q^4}({i\l \over 2}+{1\over 2})}.
\ee
It is clear that in the isotropic limit $q \to 1$ one recovers the XXX hole-hole scattering amplitude presented in section 2.

The bulk $S$-matrix, solution of the Yang-Baxter equation, for the non-critical XXZ chain is then expressed as:
\ba
{\mathbb S}(\lambda,\eta)  = {S_s(\lambda, \eta) \over a(\lambda, \eta)}\begin{pmatrix}
 a(\lambda,\eta)  & & & \cr
            &b(\lambda,\eta)&c(\eta)& \cr
            &c(\eta)&b(\lambda, \eta)&\cr
            &&&a(\lambda,\eta)
\end{pmatrix} \, ,
\label{Snc}
\ea
where we now define
\be
a(\lambda, \eta) = \sin (\eta(-\lambda +  i))\, ,  ~~~~\beta(\lambda,\eta) =-\sin( \eta\lambda)\, ,
~~~~c(\eta) = \sin(i\eta)\, .
\ee

The transmission amplitudes associated to type-I defects are also given via
\be
T^{\pm}(\hat \lambda, \eta)  =  \exp\Big [ - \sum_{k=-\infty}^{\infty}\ {1\over k}\ {e^{-2i\eta k\l}}\ \hat r_t^{\pm}(k) \Big ].
\label{ST2}
\ee
$\hat \l = \tilde \l_1- \Theta$. Recalling expressions (\ref{ST2}) and the useful identity (\ref{identnc})
we can also extract the transmission amplitudes in terms of $\Gamma_q$-functions as
\ba
&&T^+(\hat \lambda, \eta )=
{\Gamma_{q^4}(-{i\hat \l \over 2} +{3\over 4} )\ \over \Gamma_{q^4}(-{i\hat \l \over 2} +{1\over 4} )} \cr
&&T^-(\hat \lambda, \eta) =
{\Gamma_{q^4}({i\hat \l \over 2} +{1\over 4} )\ \over \Gamma_{q^4}({i\hat \l \over 2} +{3\over 4} )}.
 \label{expb1c}
\ea

Alongside the $S$-matrix (\ref{Snc}), the transmission matrices ${\mathbb T}, ~\bar {\mathbb T}$
satisfy the quadratic algebra (\ref{rttb}). Thus the corresponding transmission matrices turn out to be of the familiar form:
\ba
{\mathbb T}(\hat \l, \eta)= {e^{-i\eta \l }\ T^+(\hat \l, \eta) \over e^{-i\eta \hat \l}q^{1\over 2} - e^{i\eta \l}q^{-{1\over 2}}}
 \begin{pmatrix}
 e^{-i \eta \hat \lambda}q V -e^{i \eta \hat \lambda} q^{-1} V^{-1}  &  a^{\dag} \cr
 a &  -e^{i\eta \hat \lambda}q^{-1}V
\end{pmatrix}, \
\ea
and
\ba
\bar {\mathbb T}(\hat \lambda, \eta) = -  q^{{1\over 2}}\ T^-(\hat \l, \eta) \begin{pmatrix}
- e^{-i\eta \hat \lambda} V  &  -a^{\dag} \cr
- a &  e^{i\eta \hat \lambda} V-e^{-i \eta\hat \lambda} V^{-1}
\end{pmatrix}. \label{T2}
\ea
Crossing and unitarity properties (\ref{uni-cross}) are also explicitly checked and verified for the transmission matrices above.

Let us also briefly derive the transmission matrices of type-II associated to the spin $S$-representation of ${\mathfrak U}_q(\mathfrak{sl}_2)$ in the non-critical regime of the XXZ model. The BAE in this case are given by (see e.g. \cite{doikou-karaiskos}):
\be
e_y(\lambda_i-\Theta)\ e_1^N(\lambda_i)= - \prod_{j=1}^N e_2(\lambda_i - \lambda_j), ~~~~~~y =2S.
\ee
Deriving the density of the state with one hole in the filled Fermi sea on may extract the transmission amplitude as
\ba
&& T(\hat \l, \eta) = \exp \Big [-\sum_{k=-\infty}^{\infty}\ {1\over k}\ e^{-2ik\eta \hat \l}\ \hat r_t(k)\Big ] \cr
&& \hat r_t(k) = {e^{-\eta y |k|}\over 1 + e^{-2\eta |k|}} \label{identncb}
\ea
and it is easy show via (\ref{identnc}) and (\ref{identncb}) that
\be
T(\hat \l, \eta) = {\Gamma_{q^4}(-{i\hat \l \over 2} +{\tilde S \over 2} + {1 \over 4})\  \Gamma_{q^4}({i\hat \l \over 2} +{\tilde S \over 2} + {3 \over 4})\over \Gamma_{q^4}(-{i\hat \l \over 2} +{\tilde S \over 2} + {3 \over 4})\ \Gamma_{q^4}({i\hat \l \over 2} +{\tilde S \over 2} + {1 \over 4})},
\ee
$\tilde S = S -{1\over 2}$ is the shifted spin as also computed explicitly via BAE (see also \cite{doikou-karaiskos} for a detailed discussion).
The type-II transmission matrix in the non-critical regime satisfying (\ref{rttb}) is
\be
{\mathbb T}(\hat \lambda, \eta)  = {T(\hat \l, \eta) \over \sin\Big (\eta(-\hat \l +i \tilde S +{i\over 2})\Big )}\begin{pmatrix}
\sin\Big (\eta (-\hat \lambda + iS^z +{i\over 2})\Big )  & \sin (i\eta)\ S^- \cr
\sin(i\eta)\ S^+ & \sin \Big ( \eta (-\hat \lambda -iS^z + { i\over 2}) \Big)
\end{pmatrix} \, .
\label{ldefect}
\ee
It is easy to check that in the isotropic limit $q \to 1$ one recovers the results obtained in \cite{doikou-karaiskos} for the XXX model.
This concludes our derivation of transmission matrices in the non-critical regime of the XXZ spin chain.

\section{Discussion}
Type-I integrable quantum defects have been considered in the context of the (an)isotropic Heisenberg  model. The type-I defects
are associated to the quantum harmonic oscillator algebra, and are distinctly different to type-II defects, which are associated to the $\mathfrak{sl}_2$ ($\mathfrak{U}_q(\mathfrak{sl}_2)$) algebra \cite{doikou-karaiskos}. Here via the Bethe ansatz formulation the exact transmission matrices are explicitly derived. It is worth pointing out that although the transmission amplitudes derived here in the critical XXZ  case are similar (up to a shift to the spectral parameter) to the ones found in  \cite{konle, corrigan}, the corresponding transmission matrices are different to the ones identified in \cite{konle, corrigan} in the frame of the sine-Gordon model. It is however clear that in both cases the transmission matrices satisfy variations of the quantum harmonic oscillator algebra. The findings within the XXX and the non critical XXZ spin chain are derived here for the first time.

In the present investigation we have restricted our attention to the XXX and XXZ spin chains. It would be of great interest to
extend our analysis in the case of higher rank (deformed) algebras. More precisely, the implementation and study of generic representations of the generalizations of the quantum harmonic algebras in both isotropic and anisotropic case in order to derive the corresponding transmission amplitudes would a very interesting direction to pursue. It is worth pointing out that  not much progress has been achieved so far towards this direction, especially in the isotropic case. Also, depending on the values of the coupling constant it is possible to consider the formation of bound states between the particle-like excitations and the defect. This analysis may be achieved through the investigation of the poles appearing in the overall physical factor of the transmission matrix. All the above are significant issues, which hopefully will be addressed in the near future.

\end{document}